\documentclass[preprint]{aastex}
\usepackage{epsf}

\def\linebreak{\hfil\break}

\def\
{\hfil\linebreak}

\def\degree{\ifmmode {^\circ}\else {$^\circ$}\fi}
\def\mum{\ifmmode {\rm \mu {\rm m}}\else $\rm \mu {\rm m}$\fi}
\def\arcsec{\ifmmode ^{\prime \prime}\else $^{\prime \prime}$\fi}

\def\inch{\ifmmode ^{\prime \prime}\else $^{\prime \prime}$\fi}
\def\arcmin{\ifmmode ^{\prime}\else $^{\prime}$\fi}

\def\msun{\ifmmode {\rm M_{\odot}}\else $\rm M_{\odot}$\fi}
\def\dacrit{\ifmmode {\Delta a_{\rm crit}}\else $\Delta a_{\rm crit}$\fi}
\newbox\grsign \setbox\grsign=\hbox{$>$} \newdimen\grdimen \grdimen=\ht\grsign
\newbox\simlessbox \newbox\simgreatbox
\setbox\simgreatbox=\hbox{\raise.5ex\hbox{$>$}\llap
     {\lower.5ex\hbox{$\sim$}}}\ht1=\grdimen\dp1=0pt
\setbox\simlessbox=\hbox{\raise.5ex\hbox{$<$}\llap
     {\lower.5ex\hbox{$\sim$}}}\ht2=\grdimen\dp2=0pt

\begin{document}

\title{A Hybrid N-body--Coagulation Code for Planet Formation} 
\vskip 7ex
\author{Benjamin C. Bromley}
\affil{Department of Physics, University of Utah, 
201 JFB, Salt Lake City, UT 84112} 
\email{e-mail:bromley@physics.utah.edu}
\author{Scott J. Kenyon}
\affil{Smithsonian Astrophysical Observatory,
60 Garden Street, Cambridge, MA 02138} 
\email{e-mail: skenyon@cfa.harvard.edu}

%
%
%

\begin{abstract}

We describe a hybrid algorithm to calculate the formation of
planets from an initial ensemble of planetesimals. The algorithm
uses a coagulation code to treat the growth of planetesimals
into oligarchs and explicit $N$-body calculations to follow
the evolution of oligarchs into planets. To validate the $N$-body 
portion of the algorithm, we use a battery of tests in planetary 
dynamics.  Several complete calculations of terrestrial planet
formation with the hybrid code yield good agreement with previously 
published calculations.  These results demonstrate that the 
hybrid code provides an accurate treatment of the evolution of
planetesimals into planets.

\end{abstract}



\section{INTRODUCTION}

Rocky planet formation begins as gas and dust around a young star
settle into a thin disk. The emergence of planets within this disk 
is the result of three phases of evolution (Safronov 1969;
Weidenschilling 1980; Hayashi 1981; Wetherill \& Stewart 1993; Ida \&
Makino 1993; Kokubo \& Ida 1996).  Initially, coagulation of dust
particles causes a stochastic but steady growth in particle mass.  The
few largest bodies, or planetesimals, accumulate mass most quickly,
and experience a runaway growth phase. As the largest objects clear 
out the smaller ones, the planetesimal growth is oligarchic, where 
the largest objects -- ``oligarchs'' -- become isolated from their
neighbors and grow roughly at the same rate. In a final phase of
chaotic growth, mergers of oligarchs lead to the formation of a few
terrestrial planets around Sun-like stars well within 100~Myr 
\citep[e.g.,][]{wei97,cha01,kok02,kok06,kom02}.


This scenario is bolstered by recent observations.  Most, if not all,
young late-type stars have a disk of gas and dust (Backman \& Paresce
1993; Beckwith 1999; Lada 1999).  At least some of these disks will
form large planets, similar to those detected in radial velocity
studies (e.g., Marcy \& Butler~2000). The persistence of dusty disks
around older stars suggests that smaller planets form as well. Dust 
should be ejected by radiation pressure, but on-going formation of 
rocky planets can drive a collisional cascade which continually 
produces dust from a reservoir of small planetesimals (Kenyon
\& Bromley 2001, 2002a, 2002b). Stars with such debris disks include
$\beta$~Pic (Kalas et al.~2000; Wahhaj et al.~2003), $\epsilon$~Eri
(Greaves et al.~1998), HR~4796A (Jayawardhana et al.~1998; Koerner et
al.~1998) and Vega (e.g., Koerner, Sargent \& Ostroff 2001).

Interpretations of observed dusty debris disks, and our understanding
of planet formation in general, rely heavily on numerical
calculations. Two types of tools, statistical solvers and $N$-body
codes, provide complementary information about planetary disk
evolution. The effectiveness of each tool depends on $N$, the number
of particles they can track.  Statistical methods like Safronov's
(1969) particle-in-cell formalism work well when the particles are
numerous and have small mass \citep{kok96}. Current $N$-body codes are
unable to follow planetary growth in this regime. When the mass of
individual objects gets large, binary interactions become
important. Orbit evaluations then require direct $N$-body
calculations, not the ensemble averages of particle-in-cell
approaches.  Our experience with coagulation codes (Kenyon \& Luu
1998, 1999; Kenyon \& Bromley 2001, 2002b, 2003) suggests that we can
accurately evolve objects with masses below $10^{25}$--$10^{26}$~g
using the statistical approach.  However, an $N$-body code must track
particles of heavier mass (see also Kokubo \& Ida 1996, 2002).

Massive planetesimals requiring direct $N$-body evolution are
relatively rare in typical planet formation models.  This situation is
fortunate because $N$-body codes can not accurately track large numbers
of particles for long periods of time.  Simulations performed by
Chambers (2001) and some runs reported here involve $N \sim O(100)$
particles integrated over 100~Myr.  Ida, Kokubo \& Kominami (2003) ran
larger simulations with $N=10,000$, but only for a $\sim$0.5~Myr time
period and only with specialized hardware for gravitational force
calculations.

The complementary limitations of $N$-body algorithms and coagulation
codes call for an integration of both methods into a single ``hybrid
code'' (Jewell \& Alexander 1996; Weidenschilling et al.~1997)
Here we describe an algorithm that includes both a statistical
component and a direct $N$-body part. Our goal is to run
self-consistent simulations of planet formation, tracking 
objects from micron-sized dust grains to Jupiter mass planets.  
The coagulation code, described briefly below in {\S}2, accounts 
for gas, dust, and a swarm of lower mass planetesimals. Our 
$N$-body algorithm ({\S}3), is invoked only when needed to evolve 
the largest planetesimals.  In {\S}4 we discuss the hybrid code 
itself, and focus on how the $N$-body and coagulation parts 
interact. We provide tests of this code in {\S}5 and give 
preliminary results related to the problem of terrestrial 
planet formation.

\section{THE COAGULATION CODE}

The coagulation part of our hybrid code tracks dust and planetesimals 
in multiple annuli around the central star.  Kenyon \& Luu (1998, 1999) 
and Kenyon \& Bromley (2001, 2002b, 2004a) describe the algorithms and 
include a complete set of references.  Here we briefly review this 
statistical algorithm.

The physical processes which we simulate include coagulation,
fragmentation, gas drag, Poynting-Robertson drag, and radiation
pressure.  We discretize the continuous distribution of particle
masses into mass batches, assigning an integral number of particles to
each batch. To allow better resolution of the mass spectrum for
particles evolving rapidly, adjacent batches differ in mass by a
mass-dependent factor. We dynamically adjust the number of mass
batches, as needed.  Our spatial domain is cylindrical and is divided
into a set of concentric annuli about the central star. 
The annuli may be set so that either they have equal width, or
their boundaries are equally spaced in log-radius.

Masses and particle numbers evolve according to the coagulation 
equations which include the effects of collisions, Poynting-Robertson
drag, and interaction with the gas disk.  Our collisions rates come 
from geometric cross sections of particles, augmented by a gravitational
focusing factor for larger planetesimals. Collision outcomes depend
on the planetesimal tensile strength, gravitational binding energy, and
relative velocities. Depending on these parameters, collisions result 
in mergers, fragmentation, and dust production.

We follow particle velocities statistically for each mass batch in
each annulus, tracking vertical and horizontal velocity dispersions
relative to Keplerian orbits in the central plane of the disk.  These
dispersions evolve under the influence of gas drag, Poynting-Robertson
drag, viscous stirring and dynamical friction, according to a set of
Fokker-Planck equations (Hornung, Pellat \& Barge 1985; Wetherill \&
Stewart 1993; Stewart \& Ida 2000; Ohtsuki, Stewart, \& Ida 2002).

To solve the coagulation and Fokker-Planck equations, we use a 
fourth-order Runge-Kutta method.  The algorithm strictly conserves 
mass. When objects shift from one annulus to another, velocity 
dispersions are updated to conserve kinetic energy explicitly. 
At the inner and outer boundaries of the spatial domain, the mass 
batches reflect a steady, radial flow in the debris disk.  The code 
does not enforce angular momentum conservation, nonetheless, in 
test runs of $10^6$ timesteps, angular momentum is conserved to
better than 1\%.

The coagulation code evolves the swarm of $O(10^{21})$ planetesimals 
accurately until the largest objects reach roughly the mass of Pluto. 
Because these massive objects are rare, the statistical model provides
a poor estimate of their behavior.  Although the 
coagulation code can still follow the evolution of the swarm, the 
orbits of the largest objects require explicit calculations using an 
$N$-body code, which we describe in the next section.

\section{THE $N$-BODY CODE}\label{sect:nbody}

Planetary $N$-body codes must be fast and accurate. Suitable
solvers for the equations of motion are numerous, with published
descriptions of symplectic integrators (Wisdom \& Holman 1991;
Kinoshita, Yoshida \& Nakai 1991; Saha \& Tremaine 1992), and more
general time-symmetric integrators (e.g., Quinlan \& Tremaine 1990,
Hut, Makino \& McMillan 1995).  Some integrators (Vasilev 1982; Wisdom
\& Holman 1991; Ida \& Makino 1992; Fukushima 1996; Shefer 2002) speed
up the calculations by tracking only deviations from a purely
Keplerian orbit about the central star, following the proposal of
Encke (1852).  Other improvements include adaptive timestepping,
allowing integrators to take large timesteps when forces are small,
and to expend more computational resources only when particles
experience strong forces (Aarseth 1985; McMillan \& Aarseth 1993;
Skeel \& Biesiadecki 1994; Kokubo, Yoshinaga \& Makino 1997; Duncan,
Levison \& Lee 1998).

Our algorithm is based on the Encke method. We work with non-inertial,
Keplerian frames about the central star and integrate the equations of
motion in rectilinear coordinates defined relative to the Keplerian
frames.  These coordinates have a fixed orientation, aligned with the
inertial frame of the central star.  We use an adaptive block
timestepping scheme (e.g., McMillan \& Aarseth 1993) and a sixth- or
eighth-order accurate integrator based on Richardson extrapolation. We
give details of the integrator below.  For now we briefly outline the
timestepping scheme.

At the beginning of a timestep, each particle's accelerated reference
frame is initially set to its instantaneous Keplerian orbit.  If
particles are in tight groups, their reference frames can be set to
the Keplerian orbit of their mutual center of mass.  A
friends-of-friends algorithm (Huchra \& Geller 1982, Geller \& Huchra
1983) finds any such groups with a linking parameter based on the Hill
radii of the particles.  We then integrate the equations of motion in
terms of spatial variables in the accelerated reference frames over a
single timestep of length $\Delta t$.  Next we calculate higher
resolution orbits by dividing the timestep into $m$ equal substeps. In
practice, $m = 3$ seems to work best. A comparison between the lower
and higher resolution orbits produces a check of integrator
convergence.  We have implemented several criteria for convergence,
but have found that an individual particle's total energy, taken as a
fraction $\epsilon$, of its Keplerian orbital energy, gives an
effective comparison under most circumstances. If a particle's orbit
has not converged, we simply repeat the process at even higher
time resolution.

The strategy of a block timestepping scheme is to track individual
orbits and interpolate the converged orbits if needed to evaluate 
forces on particles whose orbits have not converged. This procedure 
reduces the computational load of force calculations if the number 
of particles is large ($N \gg 10$). For a smaller number of 
particles we prefer to integrate all orbits until every orbit has
converged. Energy conservation is better in this case and there is no
significant penalty in computational load.

The time integrator is an ordinary differential equation (ODE) solver 
based on Richardson extrapolation.  We start with a low-resolution 
estimate of the particle orbits over a time interval $\Delta t$ using 
a leapfrog intergrator.  Specifically, the position $\vec{x}$ and 
velocity $\vec{v}$ of a particle at the end of a time-forward step in 
the leapfrog case is
\begin{equation}
\begin{array}{lcl}
   \vec{v}(t+\Delta t/2) & = &\vec{v}(t) + \Delta t/2 \vec{a}(t) 
   \\
   \vec{x}(t+\Delta t)& = &\vec{x}(t)+\Delta t\vec{v}(t +\Delta t/2) 
   \\
   \vec{v}(t+\Delta t)& =&\vec{v}(t+\Delta t/2)+\Delta t/2 \vec{a}(t+\Delta t) 
\end{array}
\end{equation}
where $\vec{a}$ is the acceleration, which in our case depends only on
position. We label the position at the end of this single leapfrog
timestep as $\vec{x}_{0}$, and note that it contains errors of the
order of $\Delta t^3$.  Next we divide the time interval into two
equal parts and take two successive leapfrog steps to derive a better
resolved orbital position, $\vec{x}_{1}$. We can further subdivide the
time interval to obtain a position $\vec{x}_{i}$, derived from $2^i$ 
successive leapfrog timesteps. Our final position at the end of the 
time interval $\Delta t$ is a linear combination of these results,
\begin{equation}
\label{eq:richex}
\vec{x}(t+\Delta t) = \sum_{i=0}^m c_i \vec{x}_{i} + O(\Delta t^{2m+1})\ .
\end{equation}
Fourth-, sixth-, and eighth-order methods have $m = 1$, 2 and 3,
respectively and coefficients
\begin{equation}
\begin{array}{rcll}
(c_0, c_1) & = &  (4/3, -1/3)  & 4^{th}{\rm -order} \\
(c_0, c_1, c_2) & = & (1/45, -4/9, 64/45)  & 6^{th}{\rm -order} \\
(c_0, c_1, c_2, c_3) & = & (-1/2835, 4/135, -64/135, 4096/2835) &
8^{th}{\rm-order} .
\end{array}
\end{equation}

We calculate the gravitational forces between particles directly.
At the lowest resolution timesteps, we perform the O$(N^2)$ force
evaluations.  However, if we use the block timestepping scheme then
only a few particles depend on O$(N)$ interparticle forces at high 
temporal resolution.  To simulate a larger number of bodies, we have 
an O($N\log N)$ Barnes \& Hut (1986) treecode, as described in 
Barton, Bromley \& Geller (1999). Preliminary tests suggest that this 
code becomes competitive with a direct method only when $N$ is
greater than O$(10^3)$.  An alternative would be to use specialized
hardware (e.g., Hut \& Makino 1999). Here, our particle numbers are
small and we work exclusively with the direct force solver.


For realistic planet formation models, an $N$-body algorithm must
identify merger events.  We check for mergers in a fast way, by 
assuming that any merger event which occurs during a single timestep 
can involve only two particles.  For each particle we save an index 
number corresponding to some other particle which came the closest 
during the force calculations.  This indexing reduces the number of 
calculations from $O(N^2)$ to $O(N)$. For each pair of bodies, with 
indices $i$ and $j$, we check
the relative radial velocity using the quantity $\vec{x}_{ij} \cdot
\vec{v}_{ij}$, where the vectors are relative position and velocity
respectively. If the relative radial velocity is positive at the
beginning of a timestep, then we assume no merger takes
place. Otherwise, we find the minimum separation of the pair during
the timestep, interpolating if necessary. Once we have the minimum
pair separation $s_{ij}$ and relative speed $v_{ij}$, a
merger event is identified if
\begin{equation}
\label{eq:merger}
s_{ij} \leq
\min\left[
 R_i + R_j , 
 \frac{1}{v_{ij}}\left( v_{E,i} R_i + v_{E,j} R_j \right)
\right]
\end{equation}
where $v_{E,i}$ is the escape velocity at distance $s_{ij}$ from the $i^{\rm
th}$ particle, and $R_i$ is the particle's physical radius. Thus the
merger cross-section is at least the physical cross-section,
and is larger if the relative velocity is small compared to the
escape velocity of either particle.

\section{THE HYBRID CODE}

With our coagulation code and $N$-body algorithm, we track both the 
numerous low-mass planetesimals and the relatively rare high-mass
bodies in a planetary disk.  The interactions between these two
populations generally cause the larger bodies to circularize in their
orbits, while the smaller bodies tend to get gravitationally stirred,
as reflected in an overall increase in eccentricity and inclination.
Here we describe how the coagulation and $N$-body components
work together to simulate these effects.

The multiannulus coagulation code evolves all low-mass bodies. The
model grid contains $N$ concentric annuli with widths $\delta a_j$ 
centered at heliocentric distances $a_j$.  Each annulus contains
$n(m_{jl},t$) objects of mass $m_{jl}$ with orbital eccentricity 
$e_{jl}(t)$ and inclination $i_{jl}(t)$ in $M$ mass batches. When an 
object in the coagulation code reaches a preset mass, it is `promoted' 
into the $N$-body code. To set the initial orbit for this object, we use 
the three coagulation coordinates, $a$, $e$, and $i$, and select 
random values for the longitude of periastron and the argument of 
perihelion. Because the annuli have finite width $\delta a_j$,
we set the semimajor axis of the promoted object to $a_p$ =
$a_j + (0.5 - x) \delta a_j$, where $x$ is a random number between
0 and 1.  When two or more objects within an annulus are promoted
to the $N$-body code during the same timestep, we restrict the
choices of the orbital elements to minimize orbital interactions
between the newly promoted $N$-bodies.

The coagulation code also determines the simulation timestep.  This
interval, $\Delta T$, is generally larger than the lowest resolution
timestep of the $N$-body code, so we take multiple low-resolution
substeps, $\Delta t$, within that interval. The length of a substep is
set by the orbital speed of the fastest moving body; for nearly circular 
orbits this limit is roughly 1/60$^{\rm th}$ of the orbital period.

To calculate the effects of the swarm of low-mass bodies on the $N$-bodies,
we use particle-in-a-box estimates. For an N-body with index $j$, 
mass $m_j$,
eccentricity $e_j$, inclination $i_j$, horizontal velocity $h_j$, and 
vertical velocity $v_j$, we derive the time-evolution of the orbital 
eccentricity and inclination from the Fokker-Planck formulae for the
derivatives of the horizontal and vertical velocities:

\begin{equation}
\frac{dh_{vs,j,high}^2}{dt} = \sum_{k=1}^{N} \sum_{m=1}^{M} f_{jk} C~(h_{j}^2 + h_{km}^2)~m_{km}~J_e(\beta_{km})
\end{equation}

\begin{equation}
\frac{dv_{vs,j,high}^2}{dt} = \sum_{k=1}^{N}  \sum_{m=1}^{M} f_{jk} C~(v_{j}^2 + v_{km}^2)~m_{km}~J_z(\beta_{km})
\end{equation}

\noindent
for viscous stirring in the high velocity limit and

\begin{equation}
\frac{dh_{df,j,high}^2}{dt} = \sum_{k=1}^{N} \sum_{m=1}^{M} 1.4 f_{jk} C~(m_{km}h_{km}^2 - m_{j}h_{j}^2)~H_e(\beta_{km})
\end{equation}

\begin{equation}
\frac{dv_{df,j,high}^2}{dt} = \sum_{k=1}^{k=N} \sum_{m=1}^{M} 1.4 f_{jk} C~(m_{km}v_{km}^2 - m_{j}v_{j}^2)~H_z(\beta_{km})
\end{equation}
for dynamical friction in the high velocity limit.  In these expressions, 
$\beta_{km}^2 = (i_{j}^2 + i_{km}^2)/(e_{j}^2 + e_{km}^2)$,
$C = 0.2767~A_{\Lambda}~G^2 \rho_m / (h_{j}^2 + h_{km}^2)^{3/2}$, 
and the functions $H_e$, $H_z$, $J_e$, and $J_z$ are definite 
integrals \citep{ste00,oht02}.  The subscript $k$ is the annular index
while subscript $m$ is the index of a mass batch. 
The overlap fraction $f_{jk}$ is the fraction of bodies in annulus 
$k$ that approach within 2.4 $R_H$ of the $N$-body.  
We set $\rho_m = M_m/V_m$, where $M_m$ is the total mass of bodies 
with $m_m$ in annulus $l$ and $V_l$ is the volume of annulus $l$.  
Following Stewart \& Ida (2000), we also set $A_{\Lambda} = {\rm ln} 
(\Lambda^2 + 1) $, where

\begin{equation}
\Lambda = 0.1886~(h + 1.25 v)~v / v_{H,jk}^3 ~ ,
\end{equation}

\noindent
with $h^2 = h_{j}^2 + h_{km}^2$, $v^2 = v_{j}^2 + v_{km}^2$, 
$v_{H,jk} = r_{H,jk} v_{K,jk}$, $r_{H,jk}^3 = (m_j + m_{km})/M_{\odot}$,
and $v_{K,jk} = 0.5(v_{K,j} + v_{K,k})$; the subscript
$K$ denotes a Keplerian velocity \citep[see also][]{kb02b,oht02}.

When the relative velocities of particles approach the Hill velocity, 
$v_{H,jk}$, we set

\begin{equation}
C_1 = \frac{{\rm ln} (10 \Lambda^2/\langle e^2 \rangle + 1)}{10 \Lambda^2/\langle e^2 \rangle} ,
\end{equation}

\begin{equation}
C_2 = \frac{{\rm ln}(10 \Lambda^2 \langle e^2 \rangle^{1/2} + 1)}{10 \Lambda^2 \langle e^2 \rangle^{1/2}}
\end{equation}

\begin{equation}
C_3 = \frac{{\rm ln}(10 \Lambda^2 + 1)}{10 \Lambda^2}
\end{equation}

\noindent
and use

\begin{equation}
\frac{dh_{vs,j,low}^2}{dt} = \sum_{k=1}^{k=N} \sum_{l=1}^{l=M} 73 f_{jk} C_1 r_{H,jk}^4
\end{equation}

\begin{equation}
\frac{dv_{vs,j,low}^2}{dt} = \sum_{k=1}^{k=N}  \sum_{l=1}^{l=M} f_{jk} C_2 (4 \langle i^2 \rangle + 0.2 \langle e^2 \rangle^{3/2}\langle i^2 \rangle^{1/2}) r_{H,jk}^4
\end{equation}

\noindent
for viscous stirring in the low velocity limit and

\begin{equation}
\frac{dh_{df,j,low}^2}{dt} = \sum_{k=1}^{k=N} \sum_{l=1}^{l=M} 10 f_{jk} C_3 \langle e^2 \rangle r_{H,jk}^4 / h
\end{equation}

\begin{equation}
\frac{dv_{vs,j,low}^2}{dt} = \sum_{k=1}^{k=N}  \sum_{l=1}^{l=M} 10 f_{jk} C_3 \langle i^2 \rangle r_{H,jk}^4 / h
\end{equation}

\noindent
for dynamical friction in the low velocity limit \citep{oht02}.  
In these expressions,
$C_4$ = 0.3125$\rho_m H_{kl} (a_j + a_k) V_{K,jk}^3 / (M_j + M_{km})^2$,
$H_{jk}$ is the mutual scale height, 
$\langle e^2 \rangle$ = $(e_{j}^2 + e_{km}^2)/r_{H,jk}^2$, and
$\langle i^2 \rangle$ = $(i_{j}^2 + i_{km}^2)/r_{H,jk}^2$.
The combined velocity stirring is then

\begin{equation}
\frac{dh_{vs,j}^2}{dt} = \frac{dh_{vs,j,high}^2}{dt} + \frac{dh_{vs,j,low}^2}{dt}
\end{equation}

\begin{equation}
\frac{dv_{vs,j}^2}{dt} = \frac{dv_{vs,j,high}^2}{dt} + \frac{dv_{vs,j,low}^2}{dt}
\end{equation}

\noindent
for viscous stirring and

\begin{equation}
\frac{dh_{df,j}^2}{dt} = \frac{dh_{df,j,high}^2}{dt} + \frac{dh_{df,j,low}^2}{dt}
\end{equation}

\begin{equation}
\frac{dv_{df,j}^2}{dt} = \frac{dv_{df,j,high}^2}{dt} + \frac{dv_{df,j,low}^2}{dt}
\end{equation}
for dynamical friction.

To calculate the accretion rate of planetesimals onto the $N$-bodies,
we use the standard coagulation equation:

\begin{equation}
\frac{dm_j}{dt} = \sum_{k=1}^{k=N}  \sum_{l=1}^{l=M} A_{jkm} f_{jk} n_{km} m_{km}
\end{equation}

\noindent
where $A_{jmk}$ is the normalized cross-section (see Kenyon \& Luu 1999).

To calculate the stirring of planetesimals by $N$-bodies, we calculate
the appropriate Fokker-Planck terms for viscous stirring and dynamical
friction and add these results to the long distance stirring of 
\citet{wei89}

\begin{equation}
\frac{dh_{lr,km}^2}{dt} = \sum_{j=1}^{j=N_n} \frac{G^2~m_j^2~\Delta a}{V_{K,jk}(\delta a^2 + 0.5 H_{jk}^2)^2} ~ ,
\end{equation}

\noindent
where $N_n$ is the number of $N$-bodies and $\Delta a = a_j - a_k$.

At the end of each coagulation timestep, we pass the stirring and
accretion rates for each $N$-body -- $de_j/dt$, $di_j/dt$, and
$dm_j/dt$ -- to the $N$-body code. At the end of every low resolution
timestep in the $N$-body code, we modify each particle's orbit and
mass to reflect these changes. When possible, we simply redirect the
velocity vector so that the eccentricity varies independently of
inclination and semimajor axis. The $N$-body functions return an
updated particle list to the coagulation code. New orbital positions
and masses of the large objects are then reinserted into the
coagulation grid.  The coagulation calculations proceed in the
standard fashion, except that the orbital velocities of the large-mass
objects in the grid are not evolved. With a complete circuit from the
coagulation code to the $N$-body code to the coagulation code,
$N$-bodies influence the evolution of the swarm and the swarm
influences the evolution of the $N$-bodies.

\section{TESTS}

We have published elsewhere results on the performance of the
coagulation code, as described in {\S}2. However, the $N$-body and
hybrid codes are new, and our purpose here is to validate them. In
this section we first test the $N$-body code for stability, dynamic
range, accuracy, and merger resolution, mostly following Duncan,
Levison \& Lee (1998) in the validation of their SyMBA algorithm. We
then test the hybrid code against several simulations of terrestrial
planet formation at 1 AU \citep{wei97, cha01}.

The following subsections are organized according to the type of
calculation. We test stability during long term orbit integrations of
the major planets and a ``scaled outer solar system'' where the masses
of the major planets are increased by a factor of 50 (Duncan, Levison
\& Lee 1998). Then we test dynamic range with two binary planet
configurations. The code's accuracy is also established with a test to
resolve a critical orbital separation between two planetesimals which
determines whether their orbits will cross.  In considering mergers,
we reproduce the Greenzweig \& Lissauer (1990) results for planetary
accretion rates. We also derive the integration accuracy required to
track the collision between a massive object and a small
projectile on opposing circular orbits at 1~AU. Finally, we 
simulate terrestrial planet formation, following \citet{wei97}, 
who consider the evolution of km-sized planetesimals,
and Chambers (2001), who models the collisional evolution of lunar
mass bodies.  

\subsection*{Long-term Orbit Integration}

To evaluate the stability of our adaptive integrator, we evolve the
four major planets in orbit about a stationary Sun.
Figure~\ref{fig:symplectic} shows the behavior of the outer planets'
eccentricities over a 10~Myr period.  We use two different
integrators, a sixth-order accurate symplectic integrator (Yoshida
1990) and our adaptive code, also with a sixth-order accurate ODE
solver for comparison. We run both codes at low and high resolution.
The run times of the two codes are comparable at each resolution.  As
illustrated in the figure, the low resolution adaptive code obtains a
better measure of eccentricity than its symplectic counterpart. One
reason is that the adaptive code is not subject to accumulation of
phase errors as in a symplectic algorithm.  On the other hand, the
adaptive code produces a secular drift in energy over the course of
the simulation, whereas the symplectic code does not. The fractional error in
total energy at 10~Myr is $10^{-3}$ and $10^{-5}$ for the low and
high resolution adaptive code, respectively.

\subsection*{Scaled Outer Solar System}

If the masses of each of the four major planets increase by a 
factor of 50, the outer solar system is unstable (see, for example,
Duncan \& Lissauer 1998).  To verify that our code yields the
correct timescale for this instability, we made tests similar to 
one performed by Duncan, Levison \& Lee (1998). As expected, the
semimajor axis of Jupiter's orbit shrinks by a modest 
fraction of the original semimajor axis, while Saturn is ejected 
on a timescale of 1,000~years. As Saturn leaves the Solar System, 
its orbit crosses and excites the orbits of Uranus and Neptune. 
An increase in the orbital eccentricity of Uranus or Neptune leads
to a second ejection, as seen in Figure~\ref{fig:scalsolsys}.

\subsection*{Binaries}  

Duncan, Levison \& Lee (1998) note that bound binary planets whose 
center of mass revolves around a star provide a strong test of the
ability of an $N$-body integrator to handle a wide range of dynamical 
timescales.  They set up a pair of Jupiter-mass planets with a center 
of mass  on a circular orbit at 1~AU. The planets orbit each other with 
an initial binary semimajor axis of 0.0125~AU and binary eccentricity 
of 0.6. Their simulation time was 100~yr, covering about 3,200 binary 
orbits. 

We reproduce the Duncan, Levison \& Lee (1998) test. We can limit total 
energy errors to within a part in 10 million, while keeping the wall 
time well below a minute on a 1.7~GHz AMD Athlon processor. 
Figure~\ref{fig:binary} illustrates changes in the binary semimajor 
axis.

The figure also demonstrates a test of our block timestepper.  If a 
small particle is in a tightly bound orbit around a much more massive 
object, then the position of the massive particle is interpolated in 
the force calculations for the small body.  Our test simulates the 
orbit of a massless ``spy satellite'' on a polar orbit about the Earth.
Figure~\ref{fig:binary} shows that the distance of the satellite from 
the center of the Earth varies by about 100~meters during an orbit. 
There is no significant drift in its mean altitude over the course 
of 100 years, corresponding to 
roughly 600,000 satellite orbits of the Earth.

\subsection*{Planetesimals}

We next test the ability of the $N$-body code to handle the evolution
of planetesimals. First, we consider an isolated pair of
planetesimals, with masses of $2\times 10^{26}$~g. These bodies start
on corotating, circular orbits near 1~AU. We specify their orbital
separation in terms of their mutual Hill radius, $R_H =
[(m_1+m_2)/3\msun]^{1/3}$.  There is a critical separation, $\dacrit =
2\sqrt{3}R_H$, which determines the onset of chaotic behavior: If the
planetesimals' initial orbital separation is smaller than $\dacrit$,
their orbits will cross, otherwise they will simply scatter to larger
orbital separations.  Figure~\ref{fig:npack} shows the evolution of
semimajor axes in cases where orbital separations are initially
smaller than $\dacrit$ (top panel) and greater than $\dacrit$ (middle
panel).  As expected, the more closely-spaced pair experiences several
orbit crossings while the more distant pair does not.  Multiple
calculations demonstrate that our code can resolve the critical
separation $\dacrit$ to within a few percent.

Figure~\ref{fig:npack} also shows results for eight planetesimals,
each with the same mass as in the preceding case, and initially on
circular, Keplerian orbits. The separation between nearest neighbors
is $1.05 \dacrit$. The orbit crossings proceed from 
larger radius inward, as illustrated anecdotally in the figure
\citep[see also][]{wei97,kb01}.

Next we use the $N$-body code to model gravitational stirring by two
$2\times 10^{26}$~g objects in a uniform disk of 805 lower mass,
$2\times 10^{24}$~g objects (Kokubo \& Ida 1995; Weidenschilling et
al.~1997; Kenyon \& Bromley 2001).  The disk is centered at 1~AU, and
is 35~$R_H$ in annular extent, where $R_H$ is the mutual Hill radius
between large and small objects.  To speed up the code's convergence
with this relatively large number of particles, we soften the
interaction potential on a scale comparable to the physical radius of
the large bodies, assuming their density is 1~g cm$^{-3}$.

Figure~\ref{fig:2plus800} illustrates the evolution of the radial
profile of $(e^2 + i^2)^{1/2}$ for the case where the two massive
bodies are separated by $10 R_H$, and are each at an orbital distance
of $5 R_H$ away from 1~AU. The initial eccentricity and inclinations
are zero for the two massive objects, and small ($i \sim 10^{-5}$,
$e\sim 10^{-7}$) for the planetesimals.  The $N$-body results here
compare well with those from the pure coagulation code (Kenyon \&
Bromley 2001), except that the $N$-body calculations produce more
structure in the profile, particularly an enchanced scattering of the
swarm away from the more massive bodies.  This difference, a
broadening of the profile as compared to the coagulation results, is
expected from previous $N$-body calculations (Kokubo \& Ida 1995).

\subsection*{Mergers}

One test of our merger algorithm is provided by Greenzweig \& Lissauer
(1990), who studied gravitational focusing by a single planet in a
field of test particles, all orbiting a 1~\msun\ star. We consider one
of their configurations, a $10^{-6}$~\msun\ planet on a circular orbit
at 1~AU and a set of test particles on orbits with eccentricity $e =
0.007$, inclination $i=0.2$~degrees, and semimajor axes distributed in
two rings between 0.977~AU and 1.023~AU. We evolve the system
through a single close encounter between the planet and each test particle,
to determine the fraction of test particles accreted by the planet.

When the planet has a physical radius of $10^5$~km, Greenzweig \&
Lissauer (1990) report an accretion fraction $f_{\rm acc}$ = 0.140. In
their validation of the SyMBA code, Duncan, Levison \& Lee (1998)
estimate $f_{\rm acc} = 0.10\pm 0.02$.  For a planetary radius of
5,200~km, the derived accretion fractions are $f_{\rm acc} = 0.009$
(Greenzweig \& Lissauer 1990) and $f_{\rm acc} = 0.008\pm 0.002$
(Duncan, Levison \& Lee 1998).  Using the merger criterion specified
in \S\ref{sect:nbody} (eq. [\ref{eq:merger}]), our results are $f_{\rm
acc} = 0.138\pm 0.002$ (Poisson errors with 50,000 test particles) for
the $10^5$~km radius planet and $f_{\rm acc} = 0.008\pm 0.001$ for the
5,200~km radius planet.

%
%

A more extreme test of the merger algorithm is to simulate the
high-speed collision between two counterrotating objects. We consider
a small (cm-size) projectile and a larger target of some specified
radius, $r$, both with a density of 1~g/cm$^3$. The two bodies
initially are at 1~AU on opposite sides of the Sun and are on
colliding circular orbits.  The code calculates trajectories with a
fixed number of timesteps over a time interval which is randomly
distributed at values just greater than 0.25~yr. We run repeated
trials and find the minimum number of low-resolution timesteps,
$n_{t95}$, required to achieve a 95\% success rate in
detecting collisions.  

Figure~\ref{fig:xsection} illustrates our results. For targets with $r
< 1,000$~km, $n_{t95}$ is determined by the errors in the
interpolation of position between the endpoints of each timestep, and
scales approximately as $n_{t95} \sim r^{-1/3}$.  In our planet
formation simulations reported here, we typically take $O(100)$
timesteps per orbit and therefore can expect to resolve high-speed
collisions between 1,000~km objects.  If gravitational interactions
between target and projectile become important, as in cases with large
bodies and slow impact speeds, then the adaptive integrator takes
high-resolution timesteps, increasing the accuracy of the merger
algorithm. This effect is illustrated in Figure~\ref{fig:xsection},
which shows a dramatic decrease in $n_{t95}$ for targets of size
100~km or larger, depending on the adaptive code's error tolerance.
In practice, our $N$-bodies typically have radii greater than 1,000~km
and have impact speeds which are more than an order of magnitude
smaller than in a counterrotating collision. Thus, our code should
accurately resolve mergers in planet formation simulations. Next,
we test this assertion explicitly.

\subsection*{Terrestrial Planet Formation}

To illustrate the behavior of the hybrid code for less idealized 
conditions, we consider calculations of terrestrial planet formation 
at 1 AU. We compare our results with two sets of published calculations.
\citet{spa91} and \citet{wei97} use a multi-annulus code which evolves 
the masses and orbital properties of planetesimals with a coagulation 
code and follows the evolution of discrete planetary embryos with 
a Monte Carlo algorithm. As in our calculations, their code allows 
interactions between the planetesimals and the embryos. They evolve 
planetesimals for 1 Myr in two configurations, 0.86--1.14 AU and 
0.5--1.5 AU.  Here, we use a grid at 0.84--1.16 AU and compare the
\citet{wei97} results with our calculations from the pure coagulation 
code and from the hybrid code.

\citet{cha01} uses a three-dimensional $N$-body code to consider the
final phase of terrestrial planet formation at 0.4--2 AU. These
calculations begin with 150--160 lunar-mass to Mars-mass embryos 
and follow the collisional evolution for 200--500 Myr. Here, we adopt
a grid at 0.4--2 AU and compare the outcomes of two pure $N$-body
calculations and many hybrid calculations with the \citet{cha01}
results.

We begin with results for $a$ = 0.84--1.16 AU. In these calculations,
the grid has 32 radial zones each containing an initial distribution 
of planetesimals with radii of 1--4 km. The planetesimals have initial 
surface density $\Sigma = 16 (a / {\rm 1 ~ AU})^{-3/2}$ g cm$^{-2}$, 
eccentricity $e_0$ = $10^{-4}$, and inclination $i_0$ = $6 \times 10^{-5}$. 
The calculations include gas drag but do not allow fragmentation. 

The initial stages of pure coagulation and hybrid calculations are
identical. In a few thousand years, objects grow to radii of 10--20 km.
After $\sim 10^4$ yr, objects reach sizes of 100--300 km. Because
the timescales for dynamical friction and viscous stirring are shorter
than the growth timescales, the largest objects have nearly circular 
orbits while the smallest objects have eccentric orbits. Runaway growth 
then yields a handful of 1000---3000 km objects. As runaway growth
proceeds, viscous stirring continues to heat up the orbits of the 
smallest objects faster than the smallest objects damp the velocities
of the largest objects. Thus, gravitational focusing factors decrease,
accretion slows, and runaway growth ends. The evolution then enters
a period of `oligarchic growth,' where all of the largest objects 
grow at roughly the same rate \mbox{\citep[][]{kok96,kok98,kok02}}.

During oligarchic growth, the growth of the largest objects in the
hybrid calculations differs from the path followed in the pure 
coagulation models. In the hybrid models, oligarchs grow slowly 
until their orbits begin to cross \citep{kok02,kom02,kb06}. Once orbits
cross, chaotic growth leads to a rapid merger rate and the formation
of several `super-oligarchs' that accrete most of the leftover
planetesimals. The super-oligarchs accrete some of the remaining
lower mass oligarchs and scatter the rest out of the grid \citep{kb06}.

In pure coagulation models, the orbits of oligarchs are not allowed
to cross. These oligarchs slowly accrete all of the planetesimals 
within their gravitational reach. Oligarchs rarely accrete other
oligarchs. After 10--100 Myr, pure coagulation models have more
lower mass oligarchs than the hybrid models. However, the largest
oligarchs are less massive than their counterparts in the hybrid
calculations.

Figures \ref{fig:3dcoag} and \ref{fig:3dwhyb} compare some results
from two calculations at specific times. Figure~8 of \citet{wei97}
shows results for a similar calculation. In every case, the mass of
the largest object is $\sim 5-6 \times 10^{24}$ g at $10^4$ yr, $\sim
3-4 \times 10^{26}$ g at $10^5$ yr, and $\sim 10^{27}$ g at $10^6$ yr.
At early times, $e$ and $i$ decline monotonically with increasing
mass. At later times, $e$ and $i$ have bimodal distributions: objects
with $m \lesssim 10^{24}$ g have large $e$ and $i$, while more massive
objects have much smaller $e$ and $i$. The orbital anisotropy, $i/e$,
follows a similar evolution, with $i/e \sim$ 0.4 at early times and
$i/e \sim$ 0.1--0.15 at late times. Near the end of the calculations,
the largest objects lie in a flattened disk, with $i/e \approx$
0.02-0.03 \citep[see also][]{wei97}.

Figure \ref{fig:evsm} illustrates the evolution of the eccentricity in
more detail. At $10^4$ yr, dynamical friction maintains low $e$ for
the largest bodies. Because planets form fastest at the inner edge of
the grid, small objects at the inner edge have larger $e$ than small
objects at the outer outer edge.  At $10^5$ yr, dynamical friction
still maintains low $e$ for the largest bodies. Because most of the
small objects have grown to 100 km radius, these objects have slightly
smaller $e$ than other objects with $m < 10^{24}$ g.  At $10^6$ yr,
objects with $m > 10^{26}$ g have low $e \sim 0.02$, while lower mass
objects have higher $e \sim 0.1$. These results compare favorably with
Figure~9 of \citet{wei97}.

We begin our version of the \citet{cha01} calculations in a grid
of 40 annuli at $a$ = 0.4--2 AU. For the hybrid calculations, the
annuli contain an initial distribution of planetesimals with radii 
of 4--15 km. The pure $N$-body model starts with 160 `moons' with
a mass of $10^{26}$ g. In both cases, the objects have $e_0$ = 
$10^{-4}$ and $i_0$ = $6 \times 10^{-5}$. To provide some contrast 
with previous calculations, we adopt an initial surface density of 
planetesimals or moons, $\Sigma = x \Sigma_0 (a / {\rm 1 ~ AU})^{-1}$ 
with $\Sigma_0$ = 8 g cm$^{-2}$ and $x$ = 0.25--2, instead of the 
usual $\Sigma \propto a^{-3/2}$. Although the final configuration of 
planets depends on the initial surface density gradient, the general 
evolution is fairly independent of $\Sigma$.

The large radial extent of these calculations allows us to illustrate 
the sensitivity of planet formation to the heliocentric distance.
Because the timescale for growth by coagulation is 
$t \propto \Sigma / P \propto a^{5/2}$, planets grow fastest
at the inner edge of the grid \citep{lis87}. In hybrid models with 
$x$ = 1, objects with radii of $\sim$ 200 km form in $\sim 10^3$ yr 
at 0.4 AU, in $\sim 10^4$ yr at 0.95 AU, and in $\sim 10^5$ yr at 2 AU.
Oligarchs with masses of $\sim 10^{26}$ g form on a timescale
\begin{equation}
t_o \approx 1.5 \times 10^5 x^{-1} (a/{\rm 1 ~ AU})^{5/2} ~. 
\end{equation}

Once large objects start to form, the transition from runaway
to oligarchic to chaotic growth proceeds in several waves 
propagating from the inner disk to the outer disk \citep{kb06}. 
As oligarchs start to form at the outer edge of the grid,
dynamical interactions between oligarchs begin at the inner edge.
Several orbit crossings lead to dynamical interactions between
all oligarchs and a rapid increase in the merger rate. A few
large oligarchs gradually accrete many of the smaller oligarchs,
leading to a configuration with several Earth-mass planets and
a few Mars-mass `leftovers.'

In the moons calculations, dynamical interactions between the
160 original oligarchs dominate the entire evolutionary sequence. 
In the first $\sim 10^4$ yr, five mergers start to produce large
oligarchs in the inner disk. The number of oligarchs declines to 
146 in $10^5$ yr, to 115 in 1 Myr, and to 55 in 10 Myr. By 10 Myr,
the largest objects in the inner disk reach masses of 0.2--0.3 
$M_{\oplus}$ and slowly accrete the remaining moons. In the outer
disk, the smaller surface density and viscous stirring lead to a 
smaller merger rate. After $\sim$ 100 Myr, only $\sim$ 20 oligarchs 
remain in the outer disk.

To compare our results with \citet{cha01}, we consider measures of the
orbital elements of final `solar systems' from several simulations. In
our direct $N$-body and hybrid calculations, the time variation of the
mass-weighted eccentricity $\bar{e}$ (Fig.~\ref{fig:ecc-evol}) follows
a standard pattern (see Fig.~5 of \citep{cha01}).  During runaway and
oligarchic growth, protoplanets stir their surroundings and $\bar{e}$
increases.  At 10-100 Myr, the orbits of oligarchs cross, which
further excites orbital eccentricity and produces sharp peaks in
$\bar{e}$. Mergers between oligarchs reduce $\bar{e}$.  Once a few
remaining planets have fairly stable orbits, $\bar{e}$ settles to
0.05--0.15.

Hybrid models yield larger variations in $\bar{e}$ than $N$-body
calculations of large objects. Because objects grow to the 
promotion mass throughout a hybrid calculation, the merger phase
lasts longer and produces more frequent resonant interactions
at later times compared to pure $N$-body models. Thus, we often
observe several peaks in the evolution of $\bar{e}$ for hybrid
models compared to a single peak in pure $N$-body calculations.

Figure \ref{fig:eccent} shows the relation between the eccentricity
and mass of the planets produced in our calculations. Our models yield
a range in mass from $\sim$ 0.01 $M_{\oplus}$ to $\sim$ 2--3
$M_{\oplus}$. For $\sim$ 30 planets with masses, $m \gtrsim$ 0.1
$M_{\oplus}$, the orbital eccentricity is not correlated with the
mass. \citet{cha01} derived a similar result for $\sim$ 50
planets. However, our calculations yield a reasonably large ensemble
of lower mass planets with $m \sim$ 0.01-0.1 $M_{\oplus}$. In the full
ensemble of planets with $m \sim$ 0.01--3 $M_{\oplus}$, there is a
small, but significant trend of decreasing eccentricity with
increasing planet mass. From the Spearman rank and Kendall's $\tau$
tests \citep{pre92}, the probability that the derived distribution of
$m(e)$ is random is $\sim 2 \times 10^{-3}$.

To compare this result to our Solar System, we restrict
the mass range to 0.05--2 $M_{\oplus}$. For our ensemble of
$\sim$ 40 planets, the Spearman rank and Kendall's $\tau$
probabilities are $\sim$ 0.1. These tests yield probabilities
of $\sim$ 0.15--0.2 for the 4 terrestrial planets in our
Solar System. We conclude that the weak correlation of
$e$ with $m$ is consistent with architecture of our Solar
System.

As a final comparison with \citet{cha01}, we calculate several
statistical parameters to characterize the final configurations 
of our models. For a system with $N$ oligarchs,
\begin{equation}
S_m = m_l / m_t ~
\end{equation}
measures the fraction of the mass in the largest object, where 
$m_l$ is the mass of the largest oligarch and $m_t$ is the 
total mass in the grid.
The orbital spacing statistic is
\begin{equation}
S_s = \frac{6}{N - 1} \left ( \frac{a_{max} - a_{min}}{a_{max} + a_{min}} \right ) \left ( \frac{3 m_c}{2 \bar{m}} \right )^{1/4} ~ ,
\end{equation}
where $\bar{m}$ is the average mass of an oligarch,
$m_c$ is the mass of the central star, 
$a_{min}$ is the minimum semimajor axis of an oligarch, and
$a_{max}$ is the maximum semimajor axis of an oligarch.
To measure the degree of orbital excitation of a planet,
\citet{las97} describes the angular momentum deficit,
the difference between the z-component of the angular momentum 
of an orbit and a circular orbit with the same semimajor axis.
\citet{cha01} generalizes this definition as the sum over all
planets,
\begin{equation}
S_d = \frac{\sum_{j=1}^N m_j ~ \sqrt{a_j} ~ [1 - \sqrt{(1 - e_j^2)} cos i_j]}{\sum_{j=1}^{N} m_j ~ \sqrt{a_j}} ~.
\end{equation}
Finally, the mass concentration statistic 
\begin{equation}
S_c = max \left ( \frac{\sum_{j=1}^N m_j}{\sum_{j=1}^{N} m_j [ {\rm log} (a/a_j)^2]} \right )
\end{equation}
measures whether mass is concentrated in a few massive objects 
(large $S_c$; as in the Earth and Venus) or many low mass objects
(small $S_c$; as in the Kuiper belt and scattered disk of the
outer solar system).

Table 1 compares results for our 14 hybrid calculations and our two
`moons' calculations with statistics for the inner Solar System. For
calculations with $x = 1$, our moons calculations yield roughly the
same outcome as \citet{cha01}. The mass of the largest object and the
statistics agree well with those in Table 1 of \citet{cha01}. Because
we began with a shallower surface density gradient, our planetary 
systems have more massive objects and more oligarchs at larger 
heliocentric distances than \citet{cha01}.

Our hybrid models for $x$ = 1 yield results similar to the moons
calculations.  Our most massive object has a mass of 
1.1--1.8 $M_{\oplus}$ and contains 40\% to 80\% of the initial
mass in the grid. The calculations typically produce 1--2 other
massive objects comparable in mass to Venus and several less
massive objects with masses more similar to Mars. For models with
longer evolution times, the number of Mars-mass objects declines 
considerably from 100 Myr to 500 Myr. The statistical parameters
agree well with the \citet{cha01} results. The planets in all of 
our model solar systems have more eccentric orbits than the planets 
in the Solar System. However, the other statistics agree rather
well with those in the Solar System, including one model that is
more concentrated than the Solar System.

The outcomes of our calculations are sensitive to the initial mass
in the grid. Models with $x$ = 0.25 produce very low mass planets,
$\sim$ 0.1 $M_{\oplus}$, compared to models with $x$ = 1--2,
$\sim$ 1--2 $M_{\oplus}$.  The lower mass models also yield more
planets on more circular orbits and have a smaller fraction of the
total mass in the largest object.

\section{CONCLUSION}

In this paper we describe a hybrid $N$-body coagulation code for
planet formation.  We put the $N$-body part through a battery of tests
to assess its performance in planetary dynamics simulations.
Validation tests of the coagulation algorithm appear elsewhere (e.g.,
Kenyon \& Bromley 2001). We provide tests of the hybrid method by
comparing it to coagulation simulations and $N$-body output
separately. We demonstrate that all three methods quantitatively
reproduce published results for the formation of rocky planets by
mergers of $O(10^{12})$ km-sized \citep{wei97} or $O(100)$ moon-sized
\citep{cha01} planetesimals.  Table~2 gives a summary of these
tests, with references to other published work.

Although our intent here is simply to describe our method, the hybrid 
code now has the potential to match observations of the Solar System 
and debris disks \citep[e.g.][]{kb04a,kb04b}.  In addition to results 
on terrestrial planets in \S5, \citet{kb06} summarizes results on
the transition from oligarchic to chaotic growth in the terrestrial
zone. \citet{kb04c} shows how a stellar encounter can produce 
Sedna-like orbits in the outer Solar System and outlines how
observations might distinguish between locally-produced and
captured Sednas.  Future papers will describe more complete results 
on planet formation in the terrestrial zone and the trans-Neptunian 
region.

We acknowledge a generous allotment, $\sim$ 20 cpu years, of computer
time on the Silicon Graphics Origin-2000 `Alhena' through funding from
the NASA Offices of Mission to Planet Earth, Aeronautics, and Space
Science.  Advice and comments from M.~Geller are greatly appreciated!
The {\it NASA} {\it Astrophysics Theory Program} supported part of 
this project through grant NAG5-13278.

\vfill

\clearpage

\begin{deluxetable}{lccccccc}
\tablecolumns{8}
\tablewidth{0pc}
\tabletypesize{\normalsize}
\tablenum{1}
\tablecaption{Statistics for Planetary Systems at 0.4--2 AU}
\tablehead{\colhead{$\Sigma/\Sigma_0$} & \colhead{$t_{evol}$ (Myr)} & 
\colhead{N} & \colhead{$M_l$ ($M_{\oplus}$)} & \colhead{$S_m$} & 
\colhead{$S_s$} & \colhead{$S_d$} & \colhead{$S_c$}}
\startdata
0.25 & 100 &  9 & 0.1 & 0.198 & 19.9 & 0.0031 & 26.4 \\
0.25 & 300 & 10 & 0.1 & 0.213 & 22.7 & 0.0029 & 24.7 \\
0.50 & 100 & 10 & 0.4 & 0.292 & 20.1 & 0.0058 & 26.1 \\
0.50 & 100 &  8 & 0.4 & 0.314 & 25.1 & 0.0094 & 21.9 \\
1.00 & 100 &  6 & 1.3 & 0.474 & 27.7 & 0.0090 & 33.4 \\
1.00 & 100 &  8 & 1.0 & 0.526 & 23.4 & 0.0118 & 45.8 \\
1.00 & 100 &  5 & 1.1 & 0.428 & 35.6 & 0.0131 & 33.4 \\
1.00 & 100 &  6 & 1.4 & 0.498 & 28.3 & 0.0093 & 27.0 \\
1.00 & 200 &  5 & 1.1 & 0.371 & 28.7 & 0.0121 & 22.4 \\
1.00 & 200 &  5 & 1.8 & 0.644 & 32.0 & 0.0499 & 52.6 \\
1.00 & 500 &  2 & 2.1 & 0.778 & 36.1 & 0.0228 &119.4 \\
2.00 & 100 &  5 & 2.4 & 0.521 & 19.3 & 0.0328 & 38.6 \\
2.00 & 100 &  6 & 2.0 & 0.348 & 24.6 & 0.0292 & 25.7 \\
\\
1.00 & 100 & 21 & 0.8 & 0.256 & 14.2 & 0.0136 & 17.1 \\
1.00 & 100 & 15 & 0.9 & 0.417 & 28.8 & 0.0119 & 25.4 \\
\\
MVEM &  & 4 & 1.0 & 0.509 & 37.7 & 0.0018 & 89.9 \\
\enddata
\end{deluxetable}
\clearpage

\begin{deluxetable}{llll}
\tablecolumns{3}
\tablewidth{0pc}
\tabletypesize{\footnotesize}
\tablenum{2}
\tablecaption{Summary of Numerical Tests}
\tablehead{\colhead{Test name} & 
\colhead{Result (this paper)} & \colhead{Result (other references)}}

\startdata

\sidehead{\em\underline{Tests of integration accuracy:}}

Giant planet orbit integration  &
Figure 1 & --- \\
Scaled Outer Solar System & Figure 2 & 
Figure 5, Duncan, Levison \& Lee (1998)\\

\cline{1-3}\\
\sidehead{\em\underline{Tests of dynamic range:}}

Binary Jupiters & Figure 3 & 
Figure~4,
Duncan, Levison \& Lee (1998) \\
Earth satellite & Figure 3 &  --- \\

\cline{1-3}\\
\sidehead{\em\underline{Planetesimal dynamics:}}

Resolving chaos & Figure 4 & --- \\

Two planetesimals in a swarm & Figure 5 &
Figure 1, Kokubo \& Ida (1995) 
\\
\ & \ &  Figure B3, Weidenschilling et al.\ (1997)\\

\cline{1-3}\\
\sidehead{\em\underline{Merger algorithm:}}

Planetary accretion
($R=10^5$~km)
&  $f_{acc} = 0.138\pm 0.002$ &
$f_{acc} = 0.140$,
   Greenzweig \& Lissauer (1990) 
\\
\ \ \ ($f_{acc}$ is the accretion fraction) 
  & \ & 
$f_{acc} = 0.10\pm0.2$,
   Duncan, Levison \& Lee (1998) 
\\
Planetary accretion ($R=5200$~km) &
   $f_{acc} = 0.008\pm 0.001$ &
   $f_{acc} = 0.009$, 
   Greenzweig \& Lissauer (1990) 
\\
\ & \ & 
 $f_{acc} = 0.008\pm0.002$,
   Duncan, Levison \& Lee (1998)\\
high-speed collisions & Figure 6 & --- \\

\cline{1-3}\\
\sidehead{\em\underline{Hybrid code:}}

Planet formation (coagulation) & Figures 7--9 & 
Figures~8 \& 9,
Weidenschilling et al. (1997) 
\\
Terrestrial planet formation & Table 1 & 
Table~1,
Chambers (2001) 
\enddata
\end{deluxetable}
\clearpage


\newpage

\epsfxsize=6.0in
\epsffile{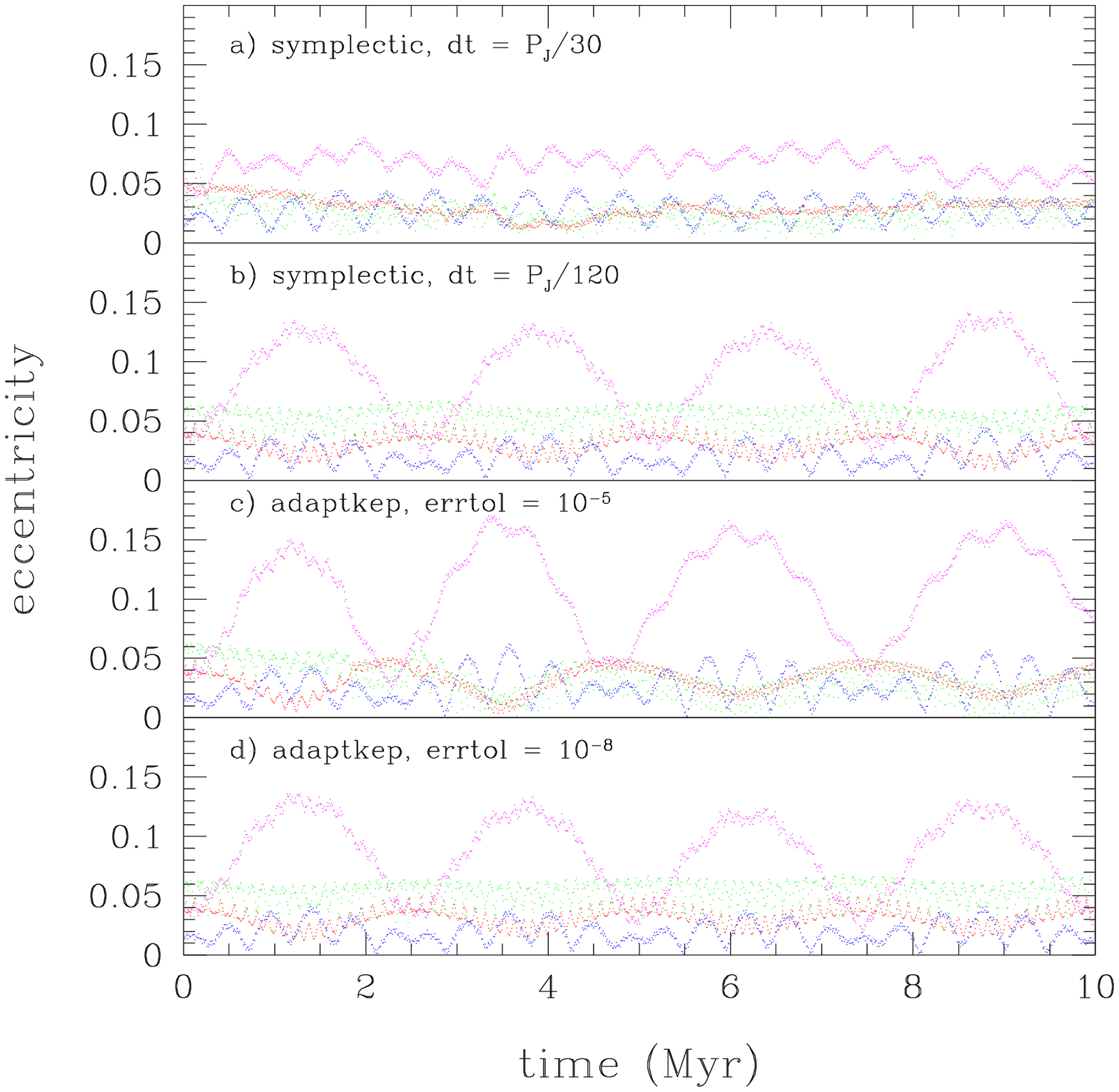} 
\figcaption[fig:symplectic] { 
\label{fig:symplectic}
The eccentricity of the major planets derive from a symplectic code 
and the adaptive code.  With a large timestep $dt$ of one thirtieth 
of the period of Jupiter's orbit ($P_J$), the symplectic integrator 
(panel a) generates phase errors---slow drifts from the true 
phase-space position of the planets. These errors are reflected
in the relatively limited range of eccentricities, as compared with a
run with smaller timesteps (panel b). The adaptive code, with a large
error tolerance, will run with large timesteps. The adaptive results 
(panel c) generate more realistic eccentricities compared to the
small-timestep symplectic code and the small-error tolerance adaptive
code (panel d).  However, the adaptive code suffers from a slow drift 
in energy. The net fractional energy errors in the adaptive runs are 
$10^{-3}$ ($10^{-5}$) for large (small) error tolerances. }

\newpage

\epsfxsize=6.0in
\epsffile{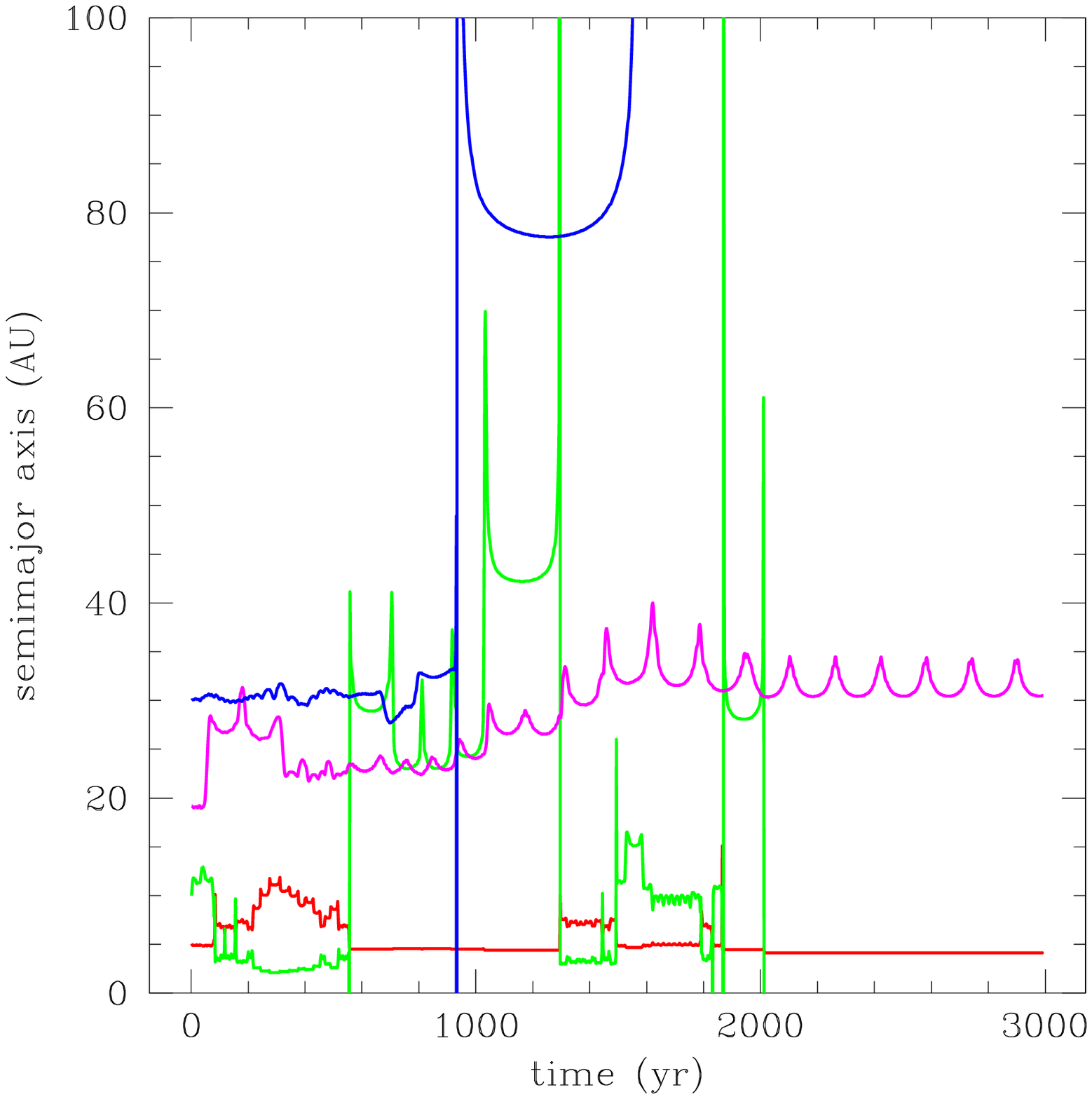} 
\figcaption { 
\label{fig:scalsolsys}
The
instantaneous Keplerian semimajor axes of the major planets, with masses
scaled by a factor of 50, as a
function of time from a simulation with our adaptive integrator. 
Saturn is ejected just after 2,000~yr; Neptune has acquired an
orbit with eccentricity of 0.92 and a semimajor axis of about 250~AU.
The overall dynamics are highly chaotic and quite sensitive to 
the outcome of the initial interaction between Jupiter and Saturn
in the first 100~yr of the simulation.
}

\newpage

\epsfxsize=6.0in
\epsffile{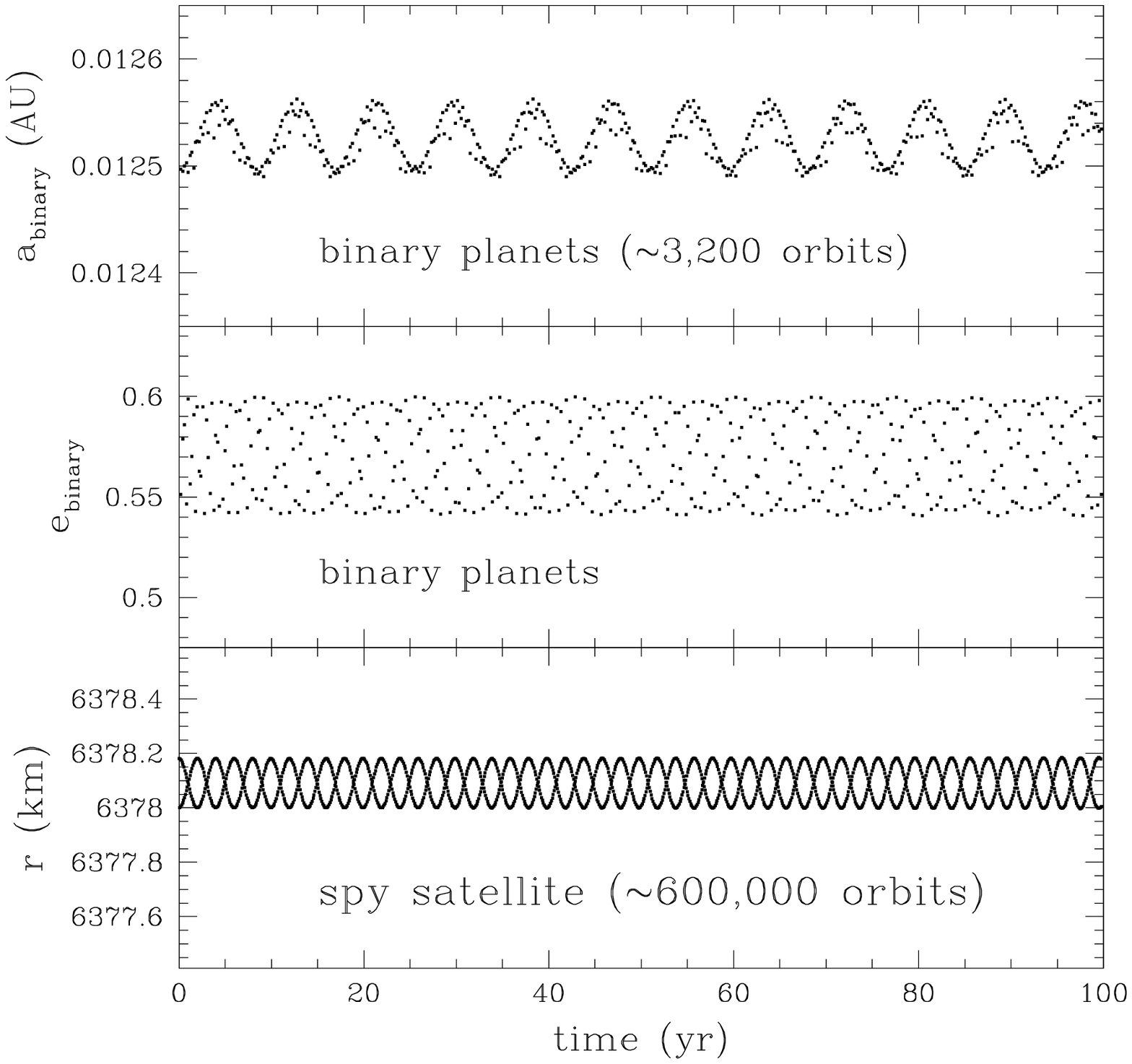} 
\figcaption[binary] { 
\label{fig:binary}
Gravitationally
bound binaries orbiting a solar-mass star.  The upper panel shows the
semimajor axis of a pair of Jupiter-mass planets as calculated in the
center-of-mass frame of the pair as they orbit the star. The binary
eccentricity is $e = 0.6$, as in Duncan, Levison \& Lee (1998). The
lower panel is the radial position of a test-particle (``spy satellite'') on
a polar orbit about the center of the Earth. In this case, we assume the
Earth is spherically symmetric. Each panel shows only a single quantity;
the appearance of multiple curves is an artifact of sampling.
}

\newpage

\epsfxsize=6.0in
\epsffile{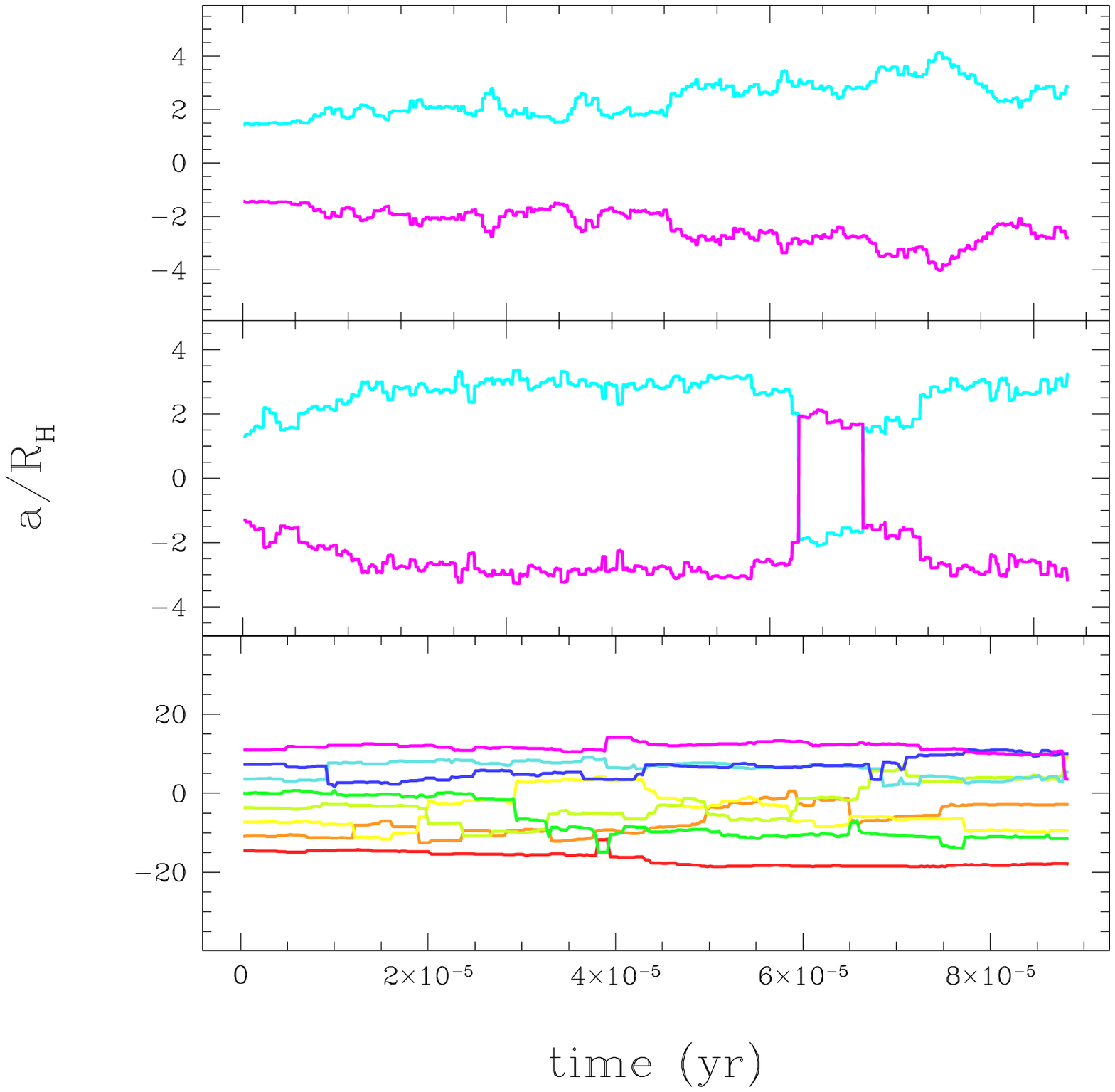} 
\figcaption[npack] {
\label{fig:npack}
Integration of planetesimal orbits. The upper two panels show the
orbital separation of a pair of $10^{26}$~g objects in units of the 
mutual Hill radius. In both cases the particles were placed on circular
orbits near 1~AU with separation of $2.1\sqrt{3} r_H$ (top) and
$1.9\sqrt{3} r_H$ (middle), where $r_H$ is the mutual Hill radius. 
When the orbital separation is less than $2\sqrt{3} r_H$, then the 
orbits will cross.  The lower panel shows interactions between 
eight particles separated by $2.1\sqrt{3} r_H$.
}

\newpage

\epsfxsize=6.0in
\epsffile{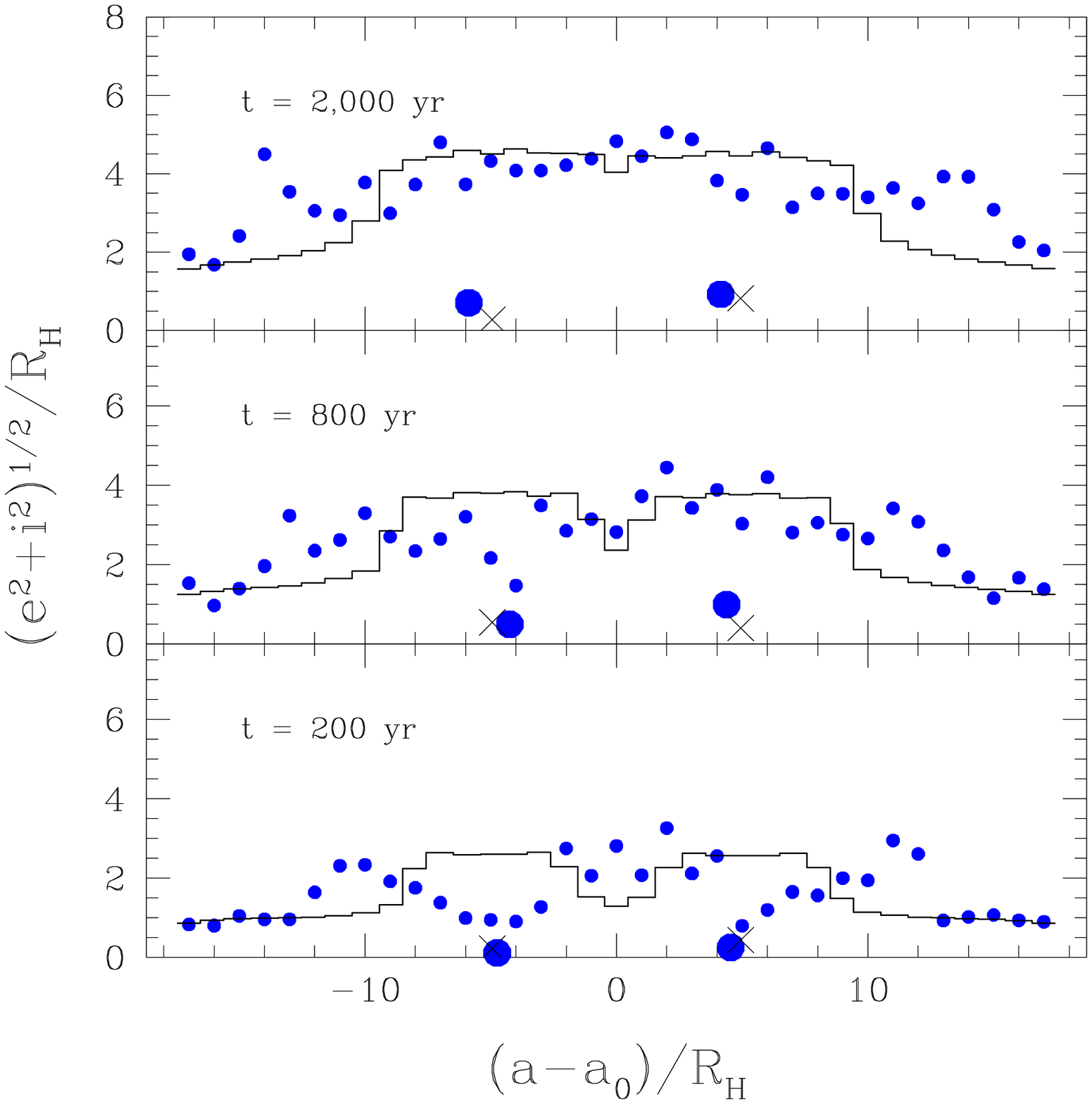} 
\figcaption[fig:2plus800]
{
\label{fig:2plus800}
Evolution of $(e^2 + i^2)^{1/2}$ in units of the Hill radius for two
$10^{26}$~g objects and a swarm of 800 $2\times 10^{24}$~g
planetesimals. The horizontal axis gives the orbital radius relative
to $a_0 = 1$~AU around a solar-type star.  The large filled circles
correspond to a pair of the heavy objects at 200 yr (lower panel), 800
yr (middle panel) and 2000 yr (top panel). The small filled circles
indicate values of $(e^2 + i^2)^{1/2}$ for the planetesimals averaged in
radial bins.  The $\times$ symbols and histograms correspond to the two 
massive objects and planetesimals in a coagulation model \citep{kb01}. 
}

\newpage

\epsfxsize=6.0in
\epsffile{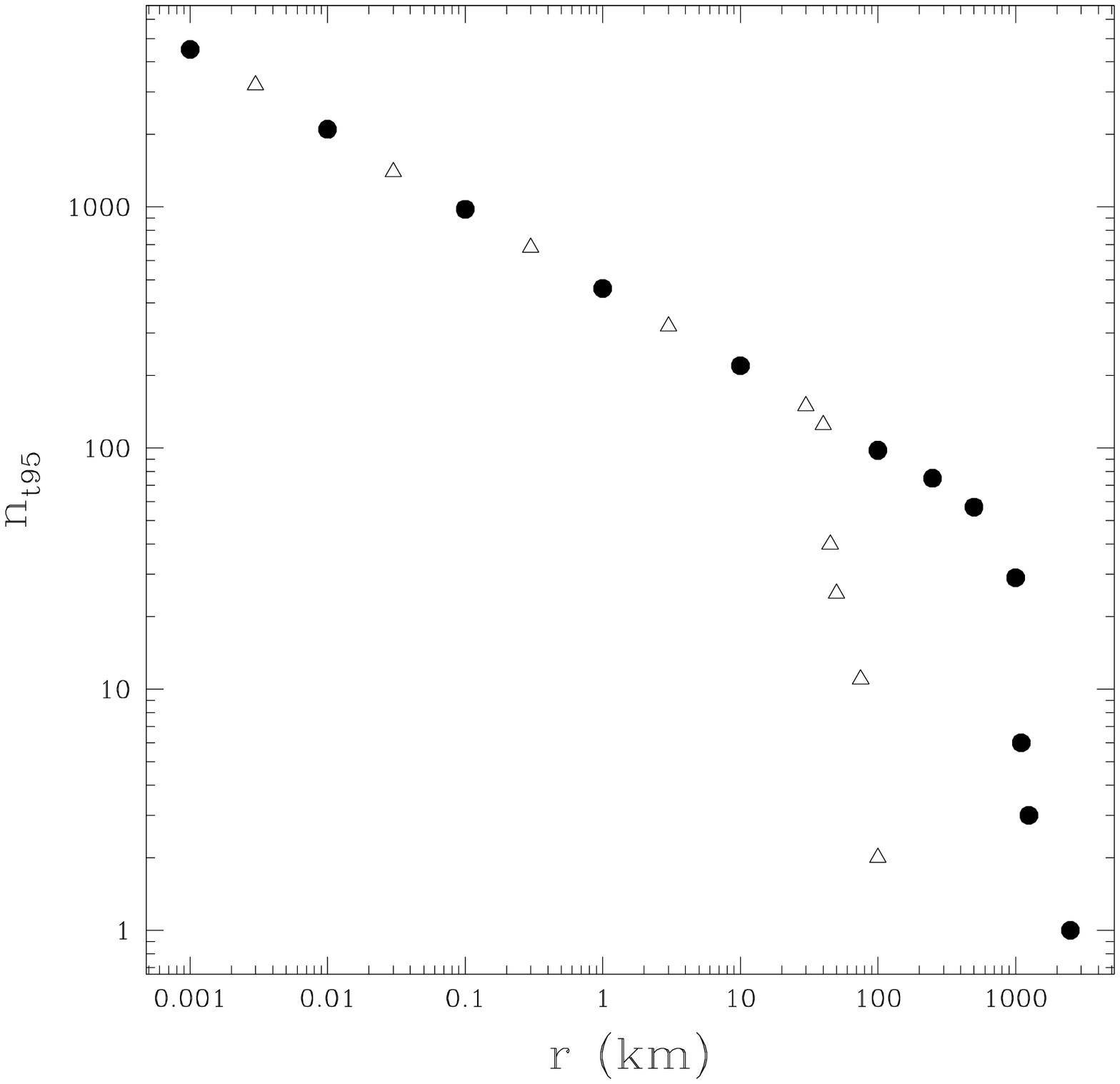} 
\figcaption[fig:xsection]
{
\label{fig:xsection}
The minimum number of timesteps leading to successful
pairwise collisions as a function of the planetesimal radius.
Specifically, $n_{t95}$ is the minimum number of low-resolution
timesteps required to resolve a merger between a point-like projectile
and a target planetesimal of radius $r$, with a success rate of 95\%.
The target and projectile are on counterrotating, circular orbits at 1~AU,
initially on opposite sides of the Sun. The error tolerance
for the adaptive code is $10^{-11}$ (filled circles) and $10^{-15}$ 
(open triangles).
The precipitous drop in $n_{t95}$ at large $r$ in both cases results from the
adaptive code's use of high-resolution timesteps as it attempts to
account for the gravitational effect of the target on the projectile.  
}

\newpage

\epsfxsize=6.0in
\epsffile{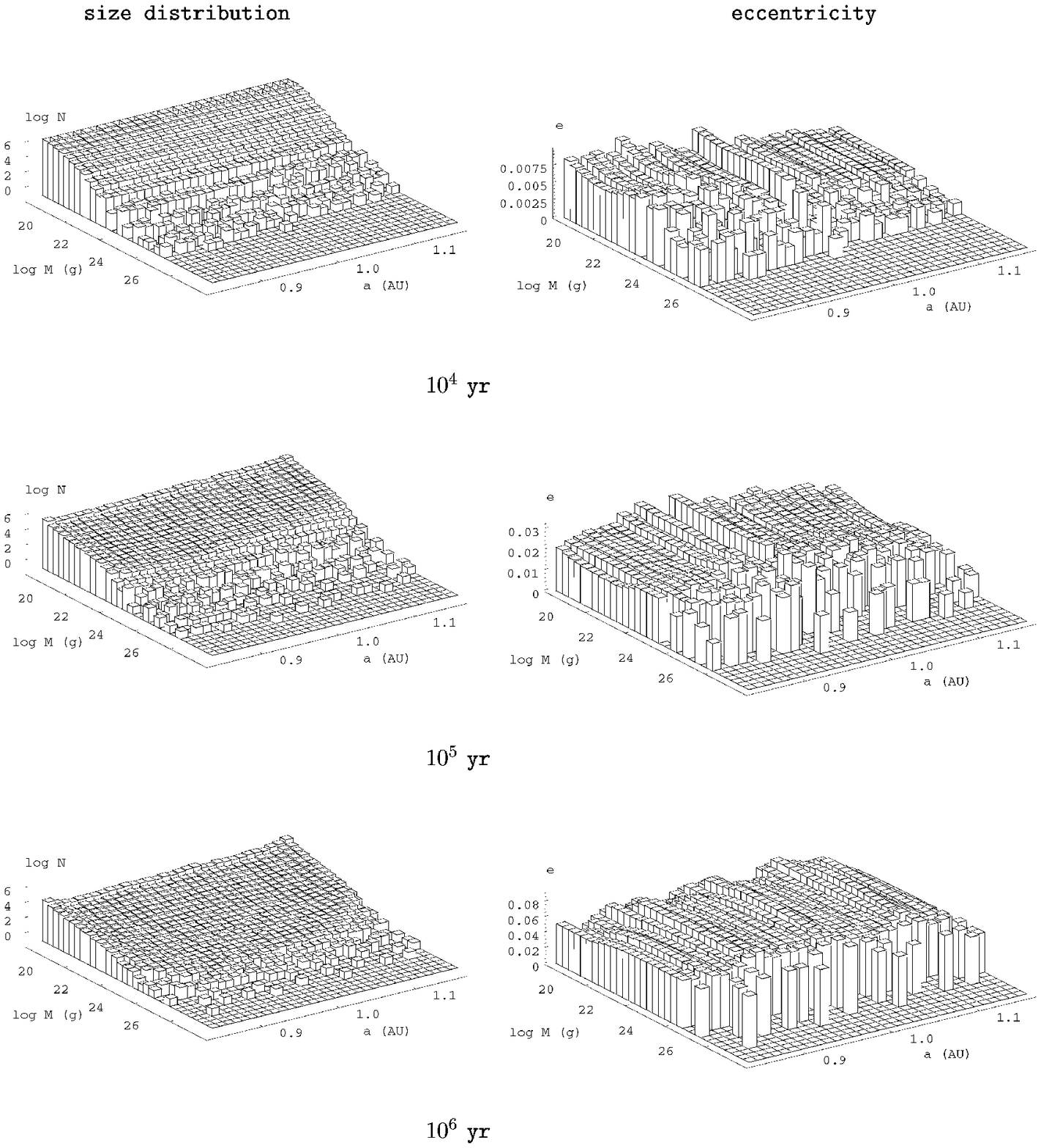} 
\figcaption[figshist3dcoag.ps]
{
\label{fig:3dcoag}
Evolution of particle number (left panels) and eccentricity (right 
panels) for planetesimals orbiting a 1 \msun~star at 1 AU.  
The 32 annuli in the grid contain planetesimals with initial radii 
of 1--4 km, initial eccentricity $e_0 = 1 \times 10^{-4}$, and a 
total mass of 0.6 $M_{\oplus}$.  
The panels show the evolution of $N(m)$ and $e$ for calculations
with the coagulation code at $t = 10^4$ yr (upper panels), 
$t = 10^5$ yr (middle panels), and $t = 10^6$ yr (lower panels).
Style adapted from Weidenschilling et al. (1997).
}

\newpage

\epsfxsize=6.0in
\epsffile{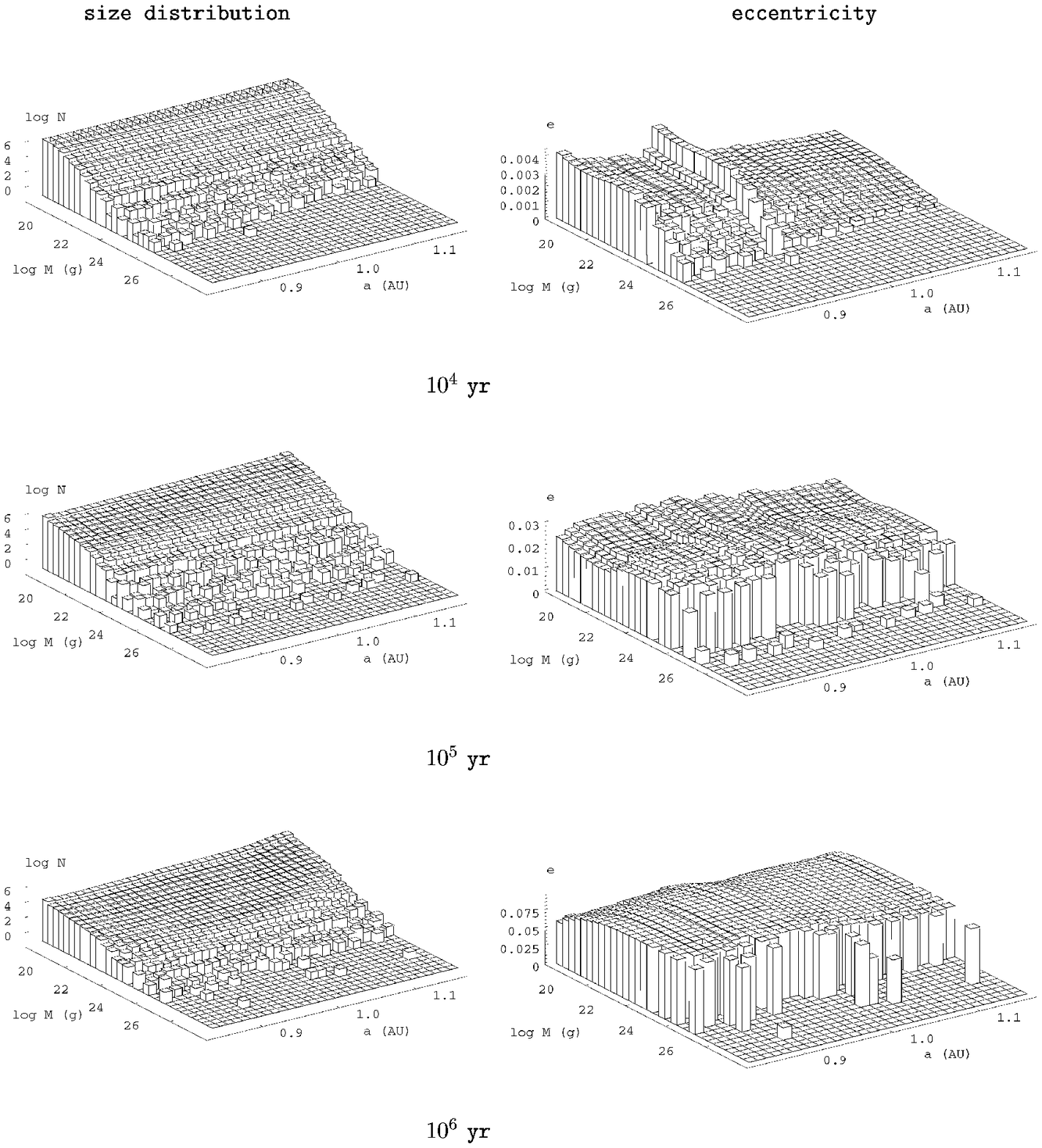} 
\figcaption[figshist3dwhyb.ps]
{
\label{fig:3dwhyb}
Evolution of particle number (left panels) and eccentricity (right 
panels) for planetesimals in orbit around a 1 \msun~star at 1 AU,
as in Figure 7, but for calculations with the hybrid code. 
}

\newpage

\epsfxsize=6.0in
\epsffile{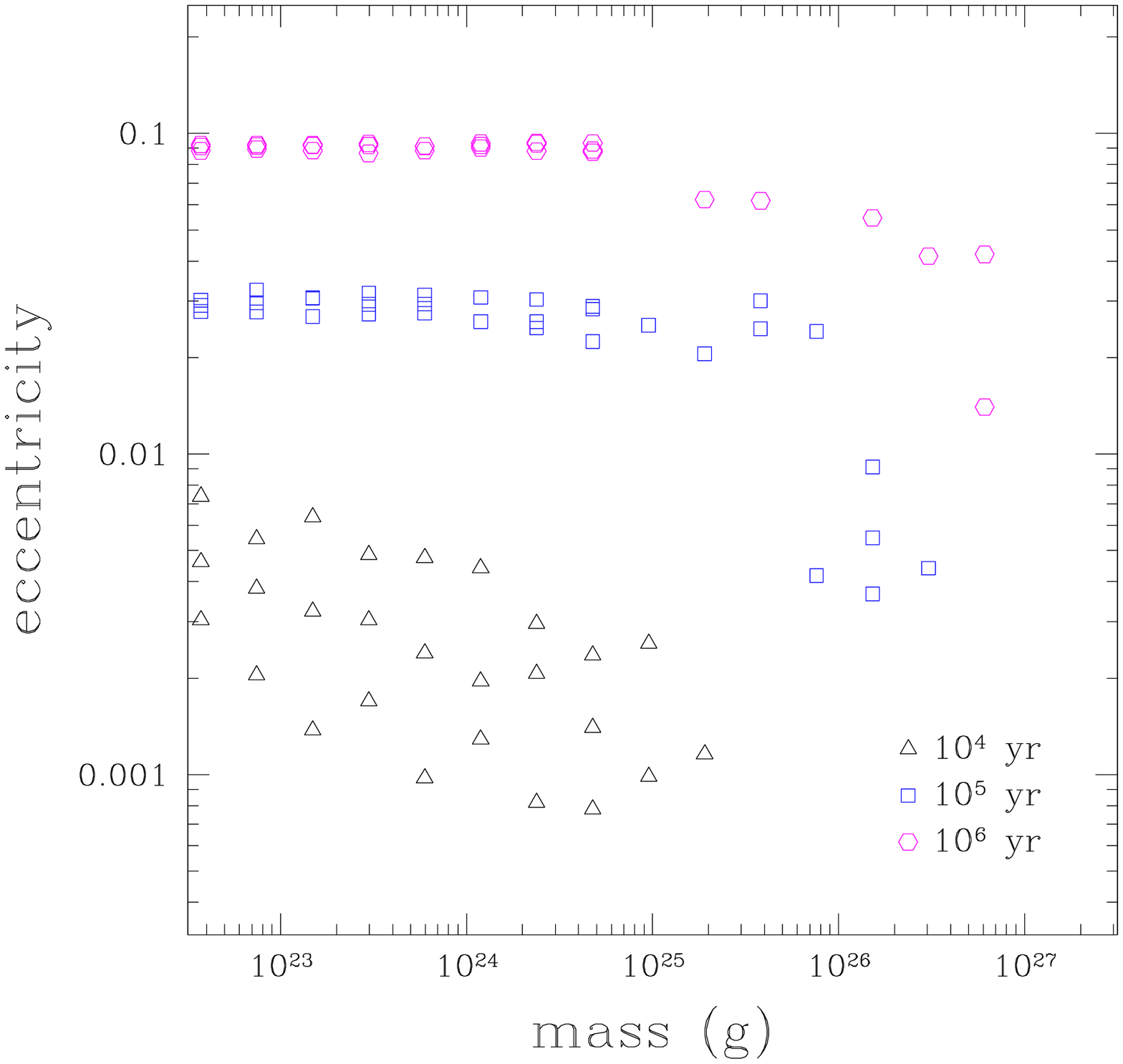} 
\figcaption[nbodyfigweidevsm.ps]
{
\label{fig:evsm}
Evolution of eccentricity as a function of mass for hybrid
calculations at 1 AU. The plots show the eccentricity at three
different times, as indicated by the legend. For smaller mass values
($m < 10^{25}$~g), averages are shown for three different annular
regions within the computational grid.  Typically, the eccentricities
are higher for objects at smaller annuli, an effect which is more
pronounced at earlier times.  For larger mass values ($m >
10^{25}$~g), the eccentricities of single objects are shown. These
results illustrate the increase in viscous stirring of smaller mass
objects with time, as well as the dynamical friction generally experienced by
the most massive bodies.  }

\newpage

\epsfxsize=6.0in
\epsffile{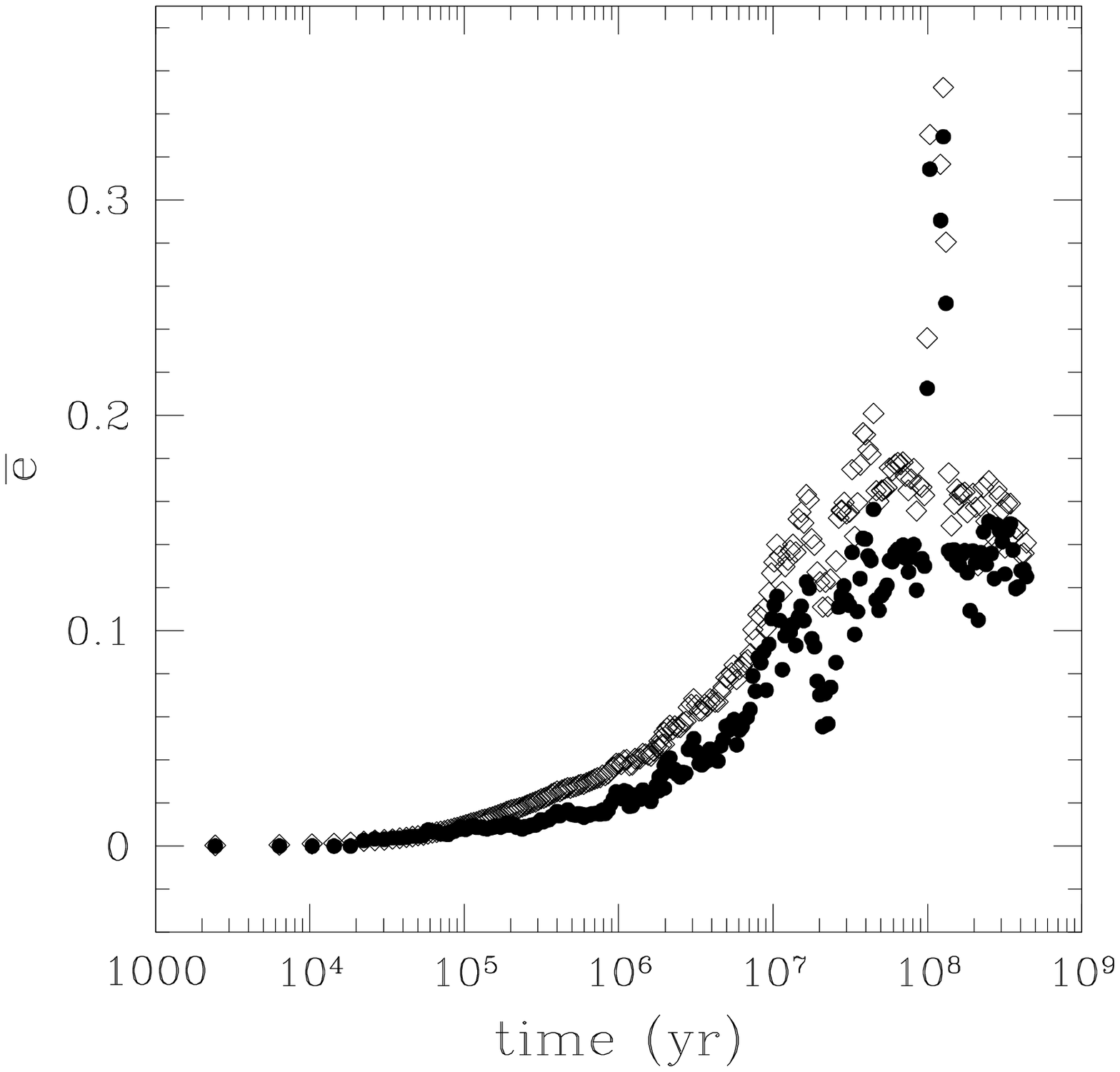} 
\figcaption[ecc-evol.eps]
{
\label{fig:ecc-evol}
Evolution of the mass-weighted eccentricity $\bar{e}$ for a 
hybrid model. The solid circles show the evolution of the 
oligarchs; the open diamonds indicate the evolution of all 
objects in the grid.  Viscous stirring and dynamical friction 
lead to a slow increase in $\bar{e}$. Strong dynamical interactions
produce abrupt rises in $\bar{e}$; mergers produce abrupt declines.
}

\newpage

\epsfxsize=6.0in
\epsffile{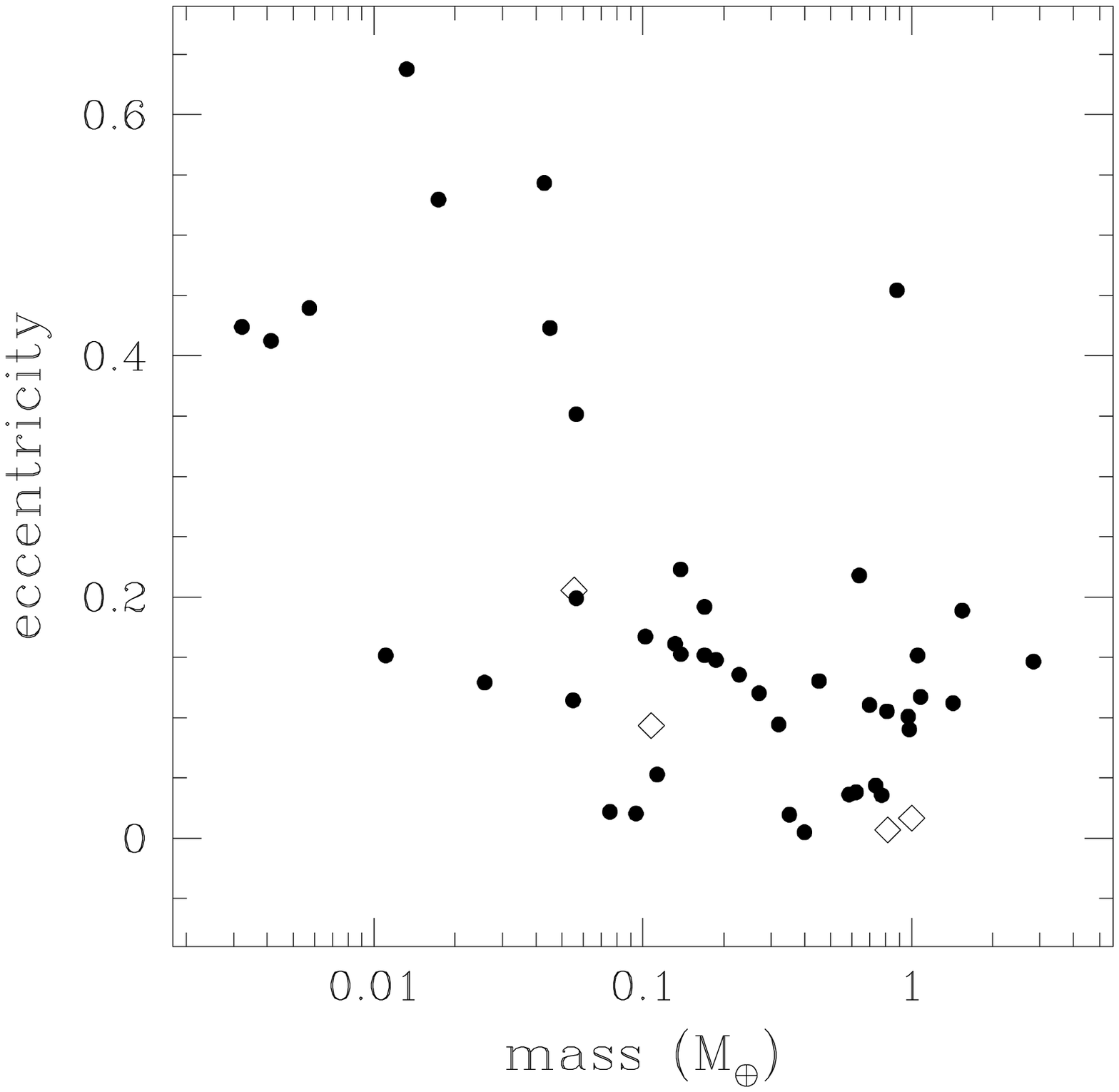} 
\figcaption[eccent.eps]
{
\label{fig:eccent}
Final eccentricity as a function of final mass for calculations 
at 0.4--2 AU. The open diamonds indicate current values for the
terrestrial planets, Mercury, Venus, Earth, and Mars. The filled
circles plot results for the ensemble of hybrid calculations 
listed in Table 1.
}

\end{document}